# Synthesis of a New Class of Reflectionless Filter Prototypes

Matthew A. Morgan - matt.morgan@nrao.edu
Tod A. Boyd - tboyd@nrao.edu

National Radio Astronomy Observatory
1180 Boxwood Estate Rd.
Charlottesville, VA. 22903

## Abstract

A design methodology and synthesis equations are described for lumped-element filter prototypes having low-pass, high-pass, band-pass, or band-stop characteristics with theoretically perfect input- and output-match at all frequencies. Such filters are a useful building block in a wide variety of systems in which the highly reactive out-of-band termination presented by a conventional filter is undesirable. The filter topology is first derived from basic principles. Then the relative merits of several implementations and tunings are compared via simulation. Finally, measured data on low-pass and band-pass filter examples are presented which illustrate the practical advantages as well as showing excellent agreement between measurement and theory.

## I. Introduction

Filters are ubiquitous components in virtually all electronic systems, from communications to radio astronomy, and arguably the fundamental principals of filter theory and optimization have been well known for the better part of a century. Nonetheless, practical filter design and implementation remains one of the most active fields of study in the electronics community today. It is therefore somewhat surprising that the role of reflectionless or absorptive filters, in which the stop-band portion of the spectrum is absorbed rather than reflected back to the source, has been largely overlooked. Although a handful of researchers have briefly visited the idea [1]-[8], it has not received the level of formal study that conventional reflective filters have.

There are many practical situations in which the reactive termination presented by a conventional filter in its stop-band adversely affects the system performance. Mixers, for example, can be extremely sensitive to the out-of-band terminations present on any of its ports, which is precisely where a filter is most likely to be in many heterodyne applications. Wideband system designers have learned to work around this problem by routinely inserting fixed attenuators in the signal path near the mixer. Similarly, high gain amplifiers, though they may be unconditionally stable in a test fixture, can easily develop instability in a packaged environment where unintended feedback is combined with a reactive out-of-band termination on its input or output. Again, a filter is very often used adjacent to the amplifier to better define the bandwidth of the system, and the stop-band impedance of the filter may need to be padded with attenuators to avoid causing stability problems.

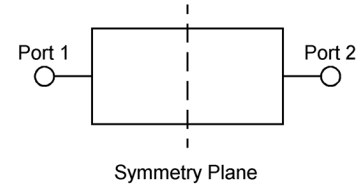

Fig. 1. A symmetric two-port network.

It is worth noting that a conventional approach to making a filter which is matched in both the pass- and stop-bands is to design a diplexer (or multiplexer) using two or more filters with complementary susceptance curves derived from singly loaded prototypes [9], and terminating all but one of these filters with a matched load. This is a fairly complex procedure, requiring at least double the number of distinct elements as a conventional filter of the same order, and would be matched on only one of the two ports.

Alternatively, one may design a directional filter using two quadrature hybrids in a balanced configuration, or by using a directional coupling structure [10]. However, quadrature hybrids with sufficient bandwidth can be difficult to design, and the intrinsically directional filter structures described do not lend themselves to high-order implementations.

To the authors' knowledge, none of the above techniques has found widespread use in industry.

In this paper we present a new class of simple reflectionless filter prototypes that do not reflect signals in their stop-bands back to the source. Indeed, the port impedances of the filters presented are theoretically constant at all frequencies. Section II describes the derivation of a circuit topology that has the desired properties. Section III discusses the optimization and comparative performance of various circuit prototypes and presents simple design equations for those found to be the most useful in practice. Section IV compares their performance to conventional Chebyshev and Butterworth filter topologies. Finally, Section V reviews the measured results of two such filters constructed to verify the circuit theory.

## II. Symmetric Reflectionless Networks

Consider an arbitrary, symmetric, two-port network, as shown in Fig. 1. (Symmetry is by no means a requirement of reflectionless filters, but it is often desirable and, as will be seen, it lends itself naturally to our design approach.) We make use of Even-Odd Mode Analysis [12]-[13], known also in an earlier form as Bartlett's Bisection Theorem [14], to analyze this network. If both ports are excited simultaneously with equal signal amplitudes and matching phase, there can be no currents crossing from one side of the symmetry plane to the other. This is called the *even-mode*. Similarly, if the two ports are excited with equal amplitudes but 180º out of phase, then all nodes that lie on the symmetry plane must have zero potential with respect to ground. This is called the *odd-mode*. One may therefore draw two single-port networks, containing only half the elements of the original two-port, and for which the nodes that lie on the symmetry plane are either open-circuited or shorted to ground. These may be called the even-mode equivalent circuit and odd-mode equivalent circuit, respectively. The scattering parameters of the original two-port network are then given as the superposition of the reflection coefficients of the even- and odd-mode equivalent circuits, as follows

$$s_{11} = s_{22} = \tfrac{1}{2}\left(\Gamma_{even} + \Gamma_{odd}\right) \tag{1}$$

$$s_{21} = s_{12} = \tfrac{1}{2}\left(\Gamma_{even} - \Gamma_{odd}\right) \tag{2}$$

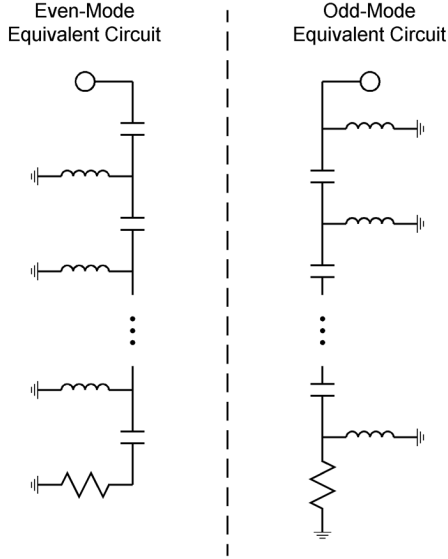

Fig. 2. Dual high-pass circuits used in the derivation of a reflectionless low-pass filter.

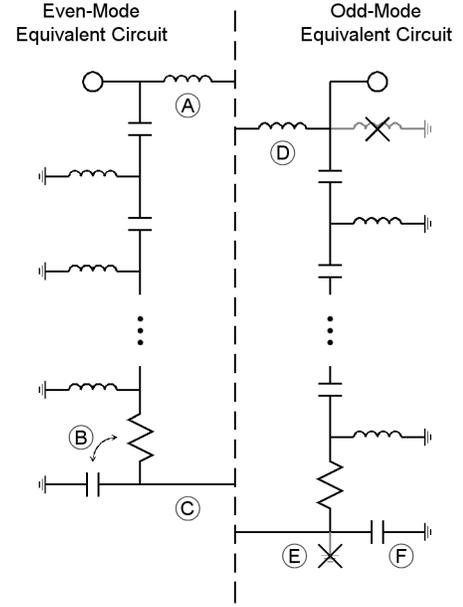

Fig. 3. Dual high-pass circuits after modification to satisfy the symmetry condition.

Thus, the condition for perfect input match, $s_{11}=0$, is derived from (1) as

$$\Gamma_{even} = -\Gamma_{odd}\,. \tag{3}$$

This is equivalent to saying that the normalized even-mode input impedance is equal to the normalized odd-mode input admittance (or vice-versa),

$$z_{even} = y_{odd} \tag{4}$$

which is satisfied if the even- and odd-mode circuits are duals of each other (i.e. inductors are replaced with capacitors, shunt elements with series elements, etc.). Further, by combining (2) and (3), we find that the transfer function of the original two-port network is given directly by the even-mode reflection coefficient,

$$s_{21} = \Gamma_{even}\,. \tag{5}$$

In the following example, we will use (4) and (5) as the basis for deriving a topology for a reflectionless low-pass filter; however, it should be noted that a very similar procedure can be followed to derive reflectionless high-pass, band-pass, and band-stop filters as well.

Equation (5) suggests that a low-pass transfer characteristic could be obtained by using a high-pass filter with one end terminated in the even-mode circuit, as shown on the left of Fig. 2. Since the transmission characteristic of this even-mode filter is high-pass, the reflection characteristic must be low-pass, and thus by (5) the completed two-port will also be low-pass. The dual of this network is then shown on the right-hand side of Fig. 2 as the odd-mode equivalent circuit.

Of course, the astute reader will realize that these two circuits cannot, as shown, be the even- and odd-mode components of the two-port network in Fig. 1, because they do not satisfy the necessary symmetry condition. However, it is possible to restore the symmetry by making topological modifications that do not affect the circuit behavior. These steps are illustrated in Fig. 3, and are described in detail

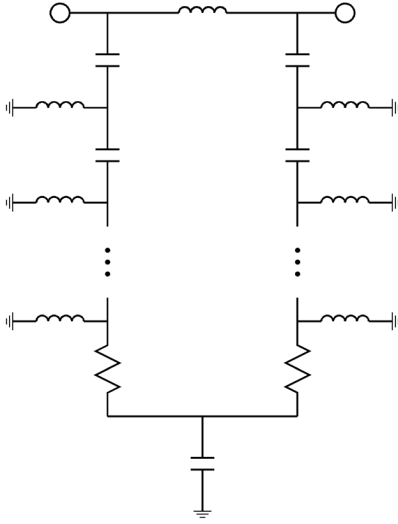

Fig. 4. Complete reflectionless low-pass filter of arbitrary order.

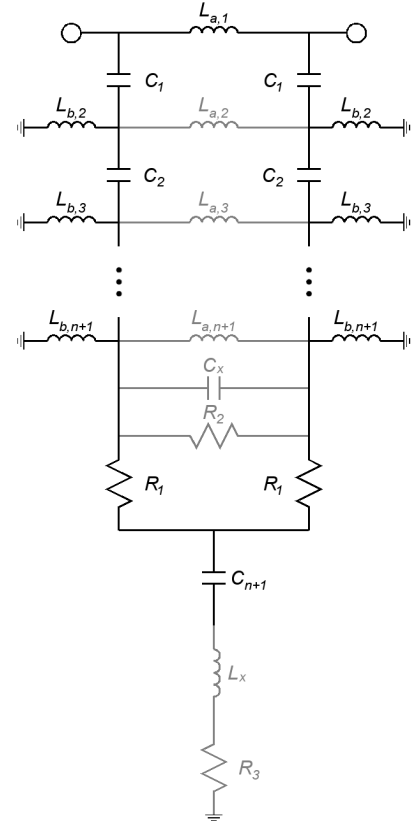

Fig. 5. Expanded $n$th-order low-pass reflectionless filter topology. The shaded elements, though providing additional degrees of freedom in component values, seem to offer no real advantage in practice.

below.

A) Add an inductor between the input node of the even-mode circuit and the symmetry plane. Since the symmetry plane equates to an open circuit in the even-mode, this inductor has no effect.

B) Reverse the order of the capacitor and the resistor, which are in series, at the end of the even-mode circuit.

C) Draw a direct connection from the node between this capacitor and resistor to the symmetry plane – again, this becomes an open circuit in the even-mode and has no effect on the circuit behavior.

D) Change the grounding of the first shunt inductor in the odd-mode circuit from absolute ground to virtual (which is defined in the odd-mode by the symmetry plane).

E) Change the grounding of the output resistor in the odd-mode circuit from absolute to virtual as well.

F) Add a capacitor between the absolute and virtual ground in the odd-mode circuit. Since both ends are grounded, it has no effect on the circuit behavior.

The final reflectionless low-pass filter topology is shown in Fig. 4.

Symmetry requires that the particular values assigned to the inductors and capacitors must be equivalent on both sides while simultaneously satisfying the duality conditions implied by Fig. 2. Additional flexibility may be obtained by adding inductors between the circuit nodes and the symmetry plane, which maintain the symmetry and have no effect on the even-mode circuit but provide an extra degree of freedom to the odd-mode circuit values. Similarly, elements of any type may be added in series with the final capacitor, and then balanced by the dual component across the symmetry plane just before the resistors. Both of these operations preserve the symmetry and allow duality constraints to be met. The resulting expanded reflectionless low-pass circuit topology is shown in Fig. 5. The values are constrained in order to satisfy duality

conditions as follows

$$C_1 = \frac{1}{Z_0^2} \cdot \frac{L_{a,1}}{2} \tag{6}$$

$$L_{b,k} = Z_0^2 \cdot C_{k-1} \tag{7}$$

$$C_k = \frac{1}{Z_0^2} \cdot \frac{L_{a,k} L_{b,k}}{L_{a,k} + 2L_{b,k}} \tag{8}$$

$$C_{n+1} = \frac{1}{Z_0^2} \cdot \frac{2L_{a,n+1} L_{b,n+1}}{L_{a,n+1} + 2L_{b,n+1}} \tag{9}$$

$$L_x = Z_0^2 C_x \tag{10}$$

$$R_2 = Z_0^2 \cdot \frac{2R_1}{R_1^2 - Z_0^2 + 2R_1 R_3} \tag{11}$$

In practice, however, the additional elements do not result in a better filter response, and can be omitted entirely, such that

$$L_{a,k} = \infty \quad k > 1 \tag{12}$$

thus reverting back to Fig. 4 as the optimal circuit topology.

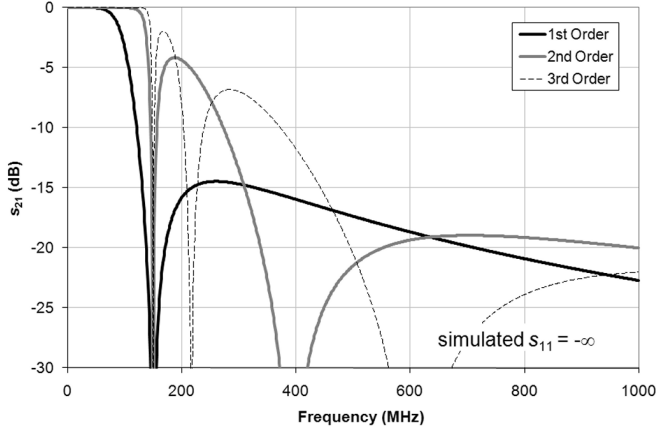

Fig. 6. Simulated transfer characteristic for 1ˢᵗ, 2ⁿᵈ, and 3ʳᵈ-order reflectionless low-pass filters. In all three cases, the first pole was located at 150 MHz.

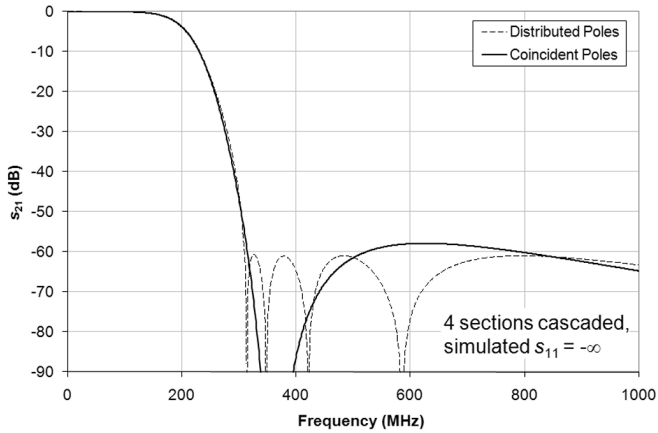

Fig. 7. Simulated comparison of cascaded reflectionless low-pass filters with 4 sections in which the poles are either coincident or distributed.

$$L_x = C_x = 0 \qquad (13)$$

$$R_2 = \infty \qquad (14)$$

$$R_3 = 0 \qquad (15)$$

## III. Cascaded-Section and Multi-Pole Filters

Simulated responses for reflectionless low-pass filters of 1ˢᵗ, 2ⁿᵈ, and 3ʳᵈ-order are shown in Fig. 6. Although higher-order filters do result in sharper cutoff characteristics, they also develop higher out-of-band peaks. The stop-band peak is just under -14 dB for the 1ˢᵗ-order filter, -4 dB for the 2ⁿᵈ-order filter, and -2 dB for the 3ʳᵈ-order filter. Unfortunately, the combined constraints of symmetry and duality derived in Section I prevent one from tuning the circuit elements to reduce these peaks. As a result, it seems unlikely that anything higher than 1ˢᵗ-order will be of much use in a real-world system. Instead, much better filter characteristics can be achieved by simply cascading 1ˢᵗ-order, single-pole filter sections. Because the filter sections are perfectly matched (in theory) there is no inherent difficulty in cascading them, and each section cascaded will provide an additional 14.5 dB of rejection in the stopband. Further, there is no need for the filter sections to be adjacent to one another. Hence, they can be distributed throughout the signal path to meet dynamic range and isolation requirements as needed for the application.

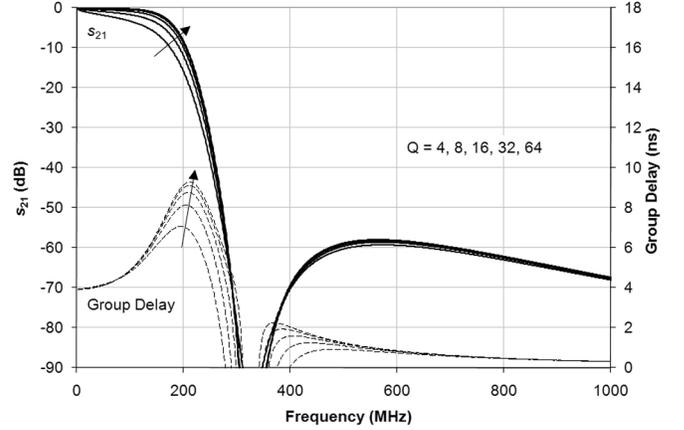

Fig. 8. Simulated gain (solid lines) and group delay (dashed lines) for a four-section, reflectionless low-pass filter, with inductor Q varying from 4 to 64.

If multiple single-pole sections are to be cascaded, the next natural question to ask is whether or not the poles should all be coincident, or whether better stop-band rejection can be achieved by distributing the poles. A comparison of these two approaches is shown in Fig. 7. Because the slope of the stop-band transfer characteristic beyond the first pole is relatively flat, there is not much advantage in spreading the poles out. With four sections in cascade, the peak stop-band rejection improves only from 58 dB to 61 dB. Since in practice the ultimate rejection at this level will depend more on the component tolerance, it is probably not worth the extra effort to tune the sections individually in this fashion.

Practical filter losses in the passband will typically be dominated by the Q of the inductors. Fig. 8 shows the simulated loss of a four-section low-pass filter with inductor Q factors of 4, 8, 16, 32, and 64, where Q was evaluated at 100 MHz. Note that the Q has almost no effect on the stop-band level, and does not impact the passband response until Q drops below about 8 using this circuit topology. Group delay is also shown on this plot, having a monotonic characteristic throughout the passband very similar to that of maximally flat filters.

As one final simplification to the design, it is noted that after substitution of (12) through (15) into (6) through (11), the first inductor, $L_{a,1}$, and the last capacitor, $C_{n+1}$, in the low-pass filter topology of Fig. 5 only differ from all the rest by a factor of two. By splitting the first inductor into two series inductors, and the last capacitor into two parallel capacitors, we arrive at a filter topology in which all the inductors, all the capacitors, and all the resistors are equal, reducing the design problem to an almost trivial selection of the pole frequency. Similar statements may be made for the high-pass, band-pass, and band-stop filters which can be derived in a manner similar to that described above. The resulting cascadable reflectionless filter prototypes are shown in Fig. 9. The frequency responses corresponding to these prototypes are shown in Fig. 10, as derived using basic, lumped-element

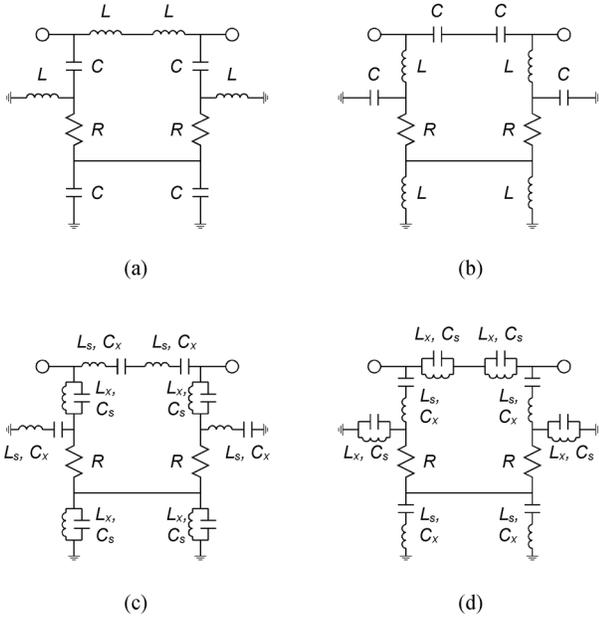

Fig. 9. First-order a) low-pass, b) high-pass, c) band-pass, and d) band-stop reflectionless filters.

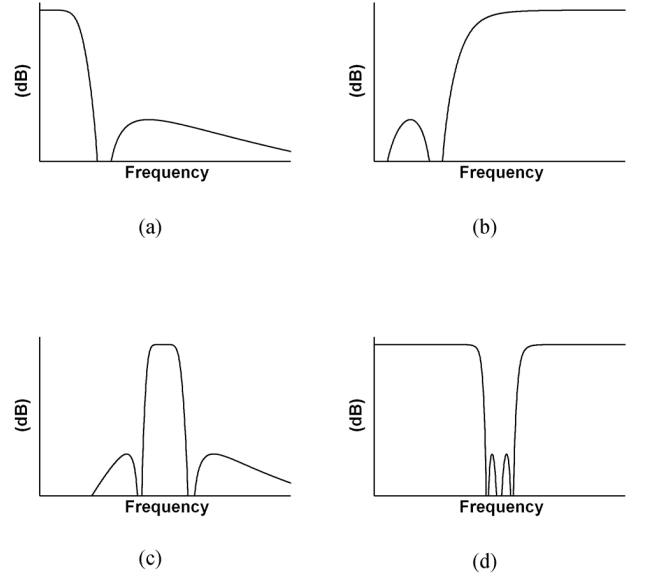

Fig. 10. Frequency response of first-order filters shown in Fig. 9. Sections may be cascaded for arbitrary levels of attenuation.

circuit analysis. To illustrate, the gain curve of the low-pass prototype in Fig. 9a is given by

$$s_{21}(f) = \frac{1 - f^2}{1 - 3f^2 + j2f(1 - f^2)} \tag{16}$$

where the frequency variable, $f$, has been normalized to the pole frequency. The stationary point at the peak of the stopband occurs at $f = \sqrt{3}$, and has a value of

$$s_{21}(\sqrt{3}) = \frac{1}{4 + j2\sqrt{3}}, \tag{17}$$

which limits the stop-band rejection per cascaded section to $10 \cdot \log(28) = 14.47$ dB.

The path the signal takes through the filters in Fig. 9 is illustrated in Fig. 11. In this figure, the elements (resistors, inductors, capacitors, and series and shunt combinations) are represented arbitrarily by rectangular impedance elements. The solid rectangles represent relatively low impedances, and the grayed-out elements represent relatively high impedances. When a signal is in the filter's pass-band, as shown in Fig. 11a, it passes directly through from one port to the other port. When a signal is in the filter's stop-band, as shown in Fig. 11b, it is blocked from passing between the ports, and instead is routed directly to the absorbing resistors in the circuit.

The design equations for these filters can be derived almost from inspection. For the low-pass and high-pass filters,

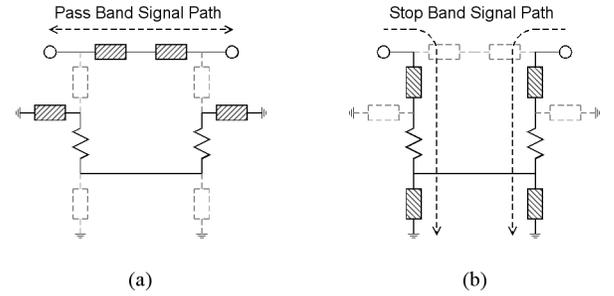

Fig. 11. Signal paths through the first order reflectionless filters in the a) pass-band and b) stop-band.

$$L = \frac{Z_0}{\omega_p} \tag{18}$$

$$C = \frac{Y_0}{\omega_p} \tag{19}$$

$$R = Z_0 \tag{20}$$

where $\omega_p$ is the pole frequency in radians/s. For band-pass and band-stop filters, which have two poles each (lower and upper), the design equations are

$$\omega_s = \omega_{p,2} - \omega_{p,1} \tag{21}$$

$$\omega_x = \frac{\omega_{p,1}\omega_{p,2}}{\omega_{p,2} - \omega_{p,1}} \tag{22}$$

$$L_s = \frac{Z_0}{\omega_s} \tag{23}$$

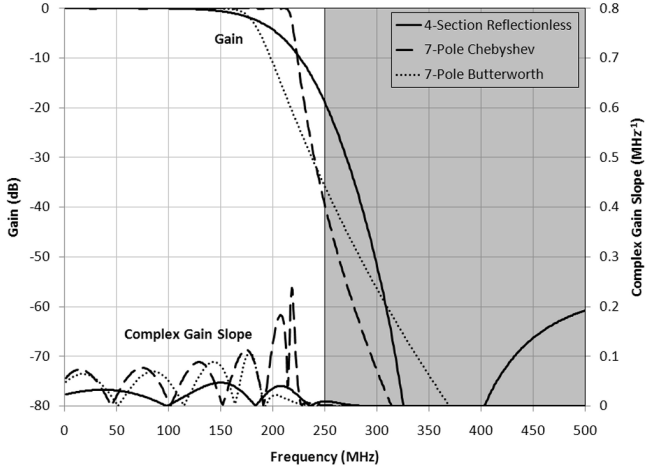

Fig. 12. Insertion loss and complex gain slope for three potential anti-aliasing filters.

$$L_x = \frac{Z_0}{\omega_x} \tag{24}$$

$$C_s = \frac{Y_0}{\omega_s} \tag{25}$$

$$C_x = \frac{Y_0}{\omega_x} \tag{26}$$

$$R = Z_0 \tag{27}$$

where $\omega_{p,1}$ and $\omega_{p,2}$ are the lower- and upper-pole frequencies, respectively.

## IV. Comparison With Conventional Filters

Although specific filter requirements will vary from one application to the next, it is useful to compare the expected performance of these filters with conventional Chebyshev or Butterworth filter topologies. As a basis for that comparison, consider the application for which these filters were originally developed – anti-aliasing in a sideband-separating downconverter for radio astronomy [15].

The sampling rate in this application was 500 MS/s. Therefore, the anti-aliasing filter was required to pass baseband signals up to 250 MHz, while rejecting aliased signals as well as LO harmonics above 250 MHz. Alias rejection was required to be at least 60 dB throughout the pass band of the filter.

Three potential filter designs are shown in Fig. 12. The first is a 4-section reflectionless low-pass filter of the type shown in Fig. 9a, the second is a 7-pole Chebyshev with 0.05 dB ripple, and the third is a 7-pole Butterworth (maximally flat) filter. All three were tuned such that the 3 dB cutoff point was as close to 250 MHz as possible, while keeping the alias rejection throughout that 3 dB bandwidth greater than 60 dB. As expected, the Chebyshev filter exhibits the sharpest cutoff and therefore has the largest bandwidth, defined by its 3 dB cutoff point, for the same level of alias rejection.

Table I: Filter Comparison: Anti-Aliasing Application (500 MS/s)

|  | 4-Section Reflectionless (this work) | 7-Pole Butterworth | 7-Pole Chebyshev |
|---|---|---|---|
| $BW_{3dB}$ | 190 MHz | 182 MHz | 219 MHz |
| Alias Suppression | 60 dB | 60 dB | 60 dB |
| Max. Inductance | 22.5 nH | 87.5 nH | 79.9 nH |
| Max. Capacitance | 9.0 pF | 34.0 pF | 26.1 pF |
| IL at 100 MHz (Q=16) | -0.8 dB | -1.4 dB | -1.6 dB |
| Complex Gain Slope | 0.05 MHz$^{-1}$ | 0.11 MHz$^{-1}$ | 0.25 MHz$^{-1}$ |

However, in this case the insertion loss is not the only parameter that defines the useful bandwidth. Of particular interest in this type of receiver is the stability of the difference in complex gain between two nominally identical anti-aliasing filters. The mismatch clearly depends on manufacturing tolerances, but in general is proportional to the complex gain slope at any given frequency,

$$\Delta G \propto \frac{d}{df} s_{21}. \tag{28}$$

In other words, the larger the gain slope, the larger the deviations that are likely to develop between two filters as a result of small manufacturing errors.

The complex gain slope for each of the three filters is plotted along the bottom of Fig. 12. Although the insertion loss of the Chebyshev filter is small over the largest bandwidth, the phase is rapidly varying throughout its passband and particularly near the cutoff, causing oscillations in the complex gain slope that are 5 times larger than that for the reflectionless filter. The second derivative of complex gain, which relates to the stability of the mismatch rather the mismatch itself, is nearly 50 times larger for the Chebyshev filter.

In fact, the filters are so stable with temperature that a calibrated sideband separating downconverter, having a matched pair of reflectionless filters in the IF circuit, was able to maintain better than 50 dB sideband isolation over a 30% RF bandwidth with the temperature varying between 28°C and 40°C [15].

Another advantage of the reflectionless filter is that it requires smaller inductors and capacitors than the more traditional designs for the same frequency band, and it can tolerate components with lower Q. This means it could be implemented using lumped elements at higher frequencies without resonances or excess insertion loss. This information is summarized in Table I.

## V. Measurements

In order to test the theory, two prototype filters were built and measured using inexpensive, discrete, surface-mount components. The first is a low-pass filter consisting of four cascaded single-pole sections shown in Fig. 9a, where all four poles are tuned to 325 MHz. The second prototype is a band-pass filter consisting of three first-order sections shown in Fig. 9c, with the lower and upper poles at 110 MHz and 310 MHz, respectively. Photographs of these prototype filters are shown in Fig. 13 and Fig. 14. Plots of their measured and simulated performance are shown in Fig. 15 and Fig. 16.

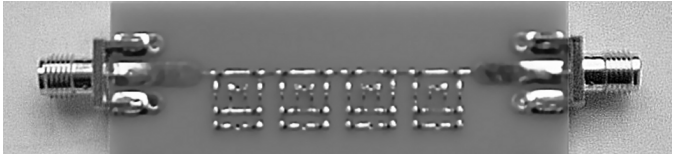

Fig. 13. Photograph of the reflectionless low-pass filter prototype.

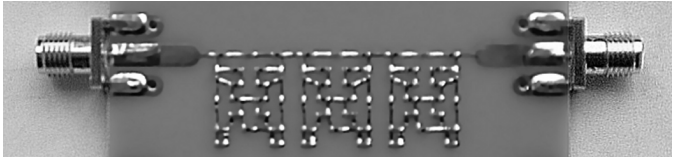

Fig. 14. Photograph of the reflectionless band-pass filter prototype.

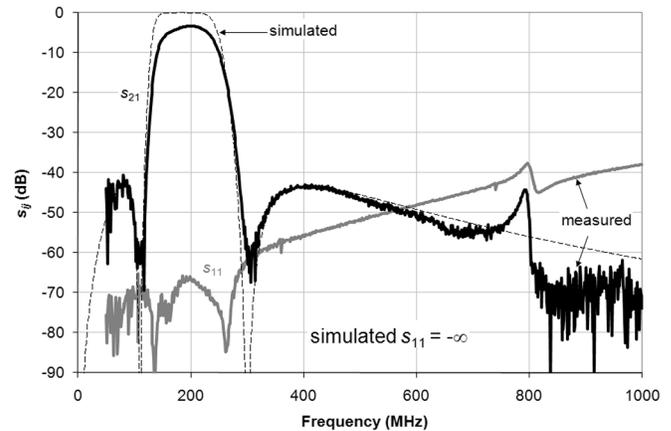

Fig. 16. Measured versus simulated performance of the reflectionless band-pass filter prototype.

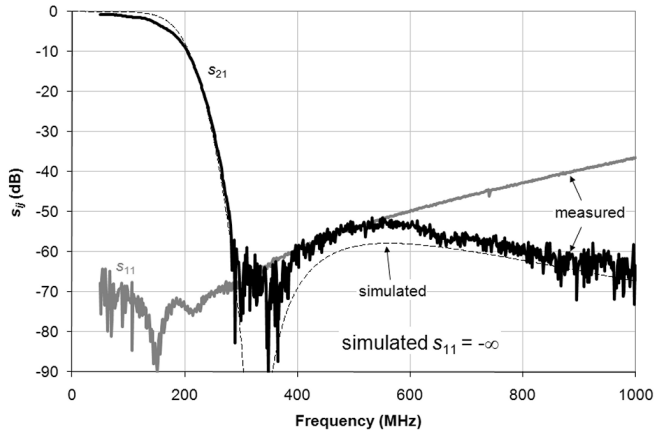

Fig. 15. Measured versus simulated performance of the reflectionless low-pass filter prototype.

The data show excellent agreement between measurement and theory, despite the fact that only ideal resistors, inductors, and capacitors were used in the simulations, and that the inductors used had only a modest Q of 8. This shows that very high-quality components, especially inductors, are not needed to implement these filters.

In simulation, the reflection coefficient is identically zero, but in practice of course, it depends on the component tolerance. The inductor and capacitor tolerances are specified to be ±2% and ±5%, respectively. However, one of the advantages of this design is that all the inductors, capacitors, resistors, series resonators, and parallel resonators have the same values throughout the filter. This allows the components to be drawn sequentially from the same lot, in which case the differential tolerance of the components relative to one another is usually much better than the absolute tolerance for a specific manufacturing process. This advantage should not be underestimated, for as the measurements show both filters achieve return loss better than 65 dB throughout their passbands, which requires a much greater level of precision than one would normally expect from components which are only specified at 5% accuracy.

The low-pass filter achieves a peak stop-band rejection of 53 dB, whereas 58 dB was simulated, and the band-pass filter achieves a level of 43 dB, matching the simulation almost exactly. There is an obvious parasitic resonance in the band-pass filter at about 800 MHz, but as this peaks at the -45 dB level, it is unlikely to cause any problem in actual use.

## VI.  Conclusion

A new class of reflectionless low-pass, high-pass, band-pass, and band-stop filter prototypes has been described. These filters function by absorbing the stop-band portion of the spectrum rather than reflecting it back to the source, which has significant advantages in some applications. The design equations for these filters are nearly-trivial, and the filters are easy and inexpensive to implement using readily available components. Measurements have shown that simple ideal-element simulations are sufficient to predict the circuit performance with astonishing accuracy. The filters are unusually stable with temperature, require smaller reactive elements than conventional filters, and have less insertion less for a given component Q.

For future work, it would be beneficial to explore modifications to the given multi-pole circuit topologies to achieve sharper cutoff characteristics, competitive with modern Chebyshev optimized filters, and to develop transmission-line implementations of these prototypes suitable for higher frequencies.

## Acknowledgment

The authors wish to thank their colleagues Tony Kerr, Sri Srikanth, Marian Pospieszalski, Shing-Kuo Pan, Rick Fisher, and John Webber for their valuable suggestions and discussion.